# Firebird Database Backup by Serialized Database Table Dump

Maurice HT Ling
Department of Zoology, The University of Melbourne, Australia

Correspondence: mauriceling@acm.org

## Abstract
This paper presents a simple data dump and load utility for Firebird databases which mimics *mysqldump* in MySQL. This utility, *fb_dump* and *fb_load*, for dumping and loading respectively, retrieves each database table using *kinterbasdb* and serializes the data using *marshal* module. This utility has two advantages over the standard Firebird database backup utility, *gbak*. Firstly, it is able to backup and restore single database tables which might help to recover corrupted databases. Secondly, the output is in text-coded format (from *marshal* module) making it more resilient than a compressed text backup, as in the case of using *gbak*.

## Categories and Subject Descriptors
    H.2.7 [**Database Administration**]: Logging and Recovery
    E.5 [**Files**]: Backup/recovery

## 1. Introduction

The only database backup and restore utility in Firebird database management system is *gbak* (chapter 38 of Borrie, 2004), which required the database to be in an uncorrupted state and backed-up into a compressed textual format known as External Data Representation (XDR) format (Srinivasan, 1995). Although some forms of database corruption can be repaired using the standard housekeeping tool, *gfix* (chapter 39 of Borrie, 2004), serious errors can still persist, preventing standard backup to be performed (page 936 of Borrie, 2004). In such cases, data within the corrupted database might be rescued by exporting the data out by third-party tools or *qli* within the standard installation.

On the other hand, MySQL's (another popular open source RDBMS) backup utility is a database dumper utility which dumps the database, table by table, as SQL insert statements. Although there are advantages with Firebird *gbak* utility (for a complete discussion of *gbak*'s capabilities, please refer to chapter 38 of Borrie, 2004), it might be useful to be able to backup or dump individual Firebird database tables into a text or text-coded format as requested by several users (Palaparambil, 2006; Piridi, 2006). A major advantage of text or text-coded format over binary compressed text format is the resilience of text format over compressed format to corruption. Corruption in text or text-coded format may result in loss of several records but corruption in compressed format may result in loss of entire tables.

This paper presents a simple data dump and load utility for Firebird databases (versions supported by kinterbasdb library which included virtually all versions of Firebird) which mimics *mysqldump* in MySQL. Each Firebird table is retrieved using *kinterbasdb* library and serialized into a text-coded file using Python's *marshal* module.

Two other third-party backup systems available for Firebird databases are nBackup (Vinkenoog, 2005) and dBak (http://www.telesiscomputing.com.au/dbk.htm) which were presented as alternatives to *gbak*. Both dBak and nBackup are comparable to the proposed method in the aspect of table-level data extraction. However, dBak relied on Borland Delphi's IBObjects; hence, unlike the proposed Python-based method, dBak is not available for Unix-based systems. Despite not being able to ascertain whether nBackup will be available for both Unix-based systems and

Microsoft systems, nBackup is only available for Firebird version 2.0 and above as a standard Firebird utility (Vinkenoog, 2005). The proposed method in this paper used *kinterbasdb* library; thus, making it available for use with all *kinterbasdb*-supported versions of Firebird which included versions lesser than 2.0. The version of Firebird used in this study is version 1.5.2. Although the proposed method in this paper lacked the ability to create incremental backups with similar ease as that of nBackup, it is able to backup multifile databases and change database owner by re-creating a clean database. Therefore, we consider our proposed method as another possible complementary Firebird database backup method (utility) to *gbak*, dBak, and nBackup.

## 2. Implementation

```
"""
Firebird database dumper which mimics MySQL database dump. The resulting files are serialized records.
Date created: October 18, 2006
Copyright (c) 2006, Maurice Ling
"""
import kinterbasdb, marshal, os, gc

db_path = 'employe2.fdb'
db_user = 'SYSDBA'
db_password = 'masterkey'
number_of_records = 500000      # number of records (tuple) in a dump file

sql_script = [
        ["cross_rate",
          "select from_currency, to_currency, conv_rate, cast(update_date as char(24)) from cross_rate",
          "insert into cross_rate (from_currency, to_currency, conv_rate, update_date) values (?,?,?,?)"]
        ]

con = kinterbasdb.connect(dsn=db_path, user=db_user, password=db_password)
cur = con.cursor()

data = {}

for table in sql_script:
    data.clear()
    gc.collect()
    data['insert_sql'] = table[2]
    print 'Dumping table: ' + table[0]
    cur.execute(table[1])
    rowcount, setcount = 0, 0
    temp = []
    for row in cur:
        temp.append(row)
        rowcount = rowcount + 1
        if (rowcount % number_of_records) == 0:
            setcount = setcount + 1
            data['data'] = temp
            temp = []
            f = open(table[0] + '.' + str(setcount) + '.dump', 'w')
            marshal.dump(data, f)
            print '%s records dumped into %s file' % (str(number_of_records), table[0] + '.' + str(setcount) + '.dump')
            f.close()
    if len(temp) > 0:
        setcount = setcount + 1
        data['data'] = temp
        temp = []
        f = open(table[0] + '.' + str(setcount) + '.dump', 'w')
        marshal.dump(data, f)
        print '%s records dumped into %s file' % (str(len(data['data'])), table[0] + '.' + str(setcount) + '.dump')
        f.close()
    print 'Total number of records in %s table: %s' % (table[0], str(rowcount))
```

```python
        print 'Dumping ' + table[0] + ' table successful.'
```

```python
"""
Firebird database loader to load dump files from fb_dump
Date created: October 18, 2006
Copyright (c) 2006, Maurice Ling
"""
import kinterbasdb, marshal, os, gc, sys

def load(f, db_path, db_user, db_password):
    con = kinterbasdb.connect(dsn=db_path, user=db_user, password=db_password)
    cur = con.cursor()
    data = marshal.load(open(f, 'r'))
    for tuple in data['data']:
        try:
            cur.execute(data['insert_sql'], tuple)
            con.commit()
        except: 'Unable to load record: %s' % tuple
    print '%s file loaded successful.' % f
    con.close()

if __name__ == "__main__":
    if len(sys.argv) < 5:
        print "Usage: python db_load.py <path_to_database> <database_user> <database_password> <dump_file>+"
    for dumpfile in sys.argv[4:]:
        load(dumpfile, sys.argv[1], sys.argv[2], sys.argv[3])
```

## 3. Evaluation

Two evaluation tests were carried out – the first evaluation demonstrated a proof of concept while the second evaluation demonstrated that the utility could be used on a larger database with reasonable performance.

In the first evaluation, a sample database was constructed using the following script as supplied as an example in the Firebird installation:

```sql
set sql dialect 1;
create database "employe2.fdb";
CREATE TABLE cross_rate(
    from_currency   VARCHAR(10) NOT NULL,
    to_currency     VARCHAR(10) NOT NULL,
    conv_rate       FLOAT NOT NULL,
    update_date     DATE,
    PRIMARY KEY (from_currency, to_currency));

INSERT INTO cross_rate VALUES ('Dollar', 'CdnDlr',  1.3273, '11/22/93');
INSERT INTO cross_rate VALUES ('Dollar', 'FFranc',  5.9193, '11/22/93');
INSERT INTO cross_rate VALUES ('Dollar', 'D-Mark',  1.7038, '11/22/93');
INSERT INTO cross_rate VALUES ('Dollar', 'Lira',    1680.0, '11/22/93');
INSERT INTO cross_rate VALUES ('Dollar', 'Yen',     108.43, '11/22/93');
INSERT INTO cross_rate VALUES ('Dollar', 'Guilder', 1.9115, '11/22/93');
INSERT INTO cross_rate VALUES ('Dollar', 'SFranc',  1.4945, '11/22/93');
INSERT INTO cross_rate VALUES ('Dollar', 'Pound',   0.67774, '11/22/93');
INSERT INTO cross_rate VALUES ('Pound',  'FFranc',  8.734, '11/22/93');
INSERT INTO cross_rate VALUES ('Pound',  'Yen',     159.99, '11/22/93');
INSERT INTO cross_rate VALUES ('Yen',    'Pound',   0.00625, '11/22/93');
INSERT INTO cross_rate VALUES ('CdnDlr', 'Dollar',  0.75341, '11/22/93');
INSERT INTO cross_rate VALUES ('CdnDlr', 'FFranc',  4.4597, '11/22/93');
```

The resulting serialized dump file is as follows:

{t^D^@^@^@data[^N^@^@^@(^D^@^@^@s^F^@^@^@Dollars^F^@^@^@CdnDlrf^Q1.3272999525070
19s^X^@^@^@1993-11-22 00:00:00.0000(^D^@^@^@s^F^@^@^@Dollars^F^@^@^@FFrancf^R5.9
193000793457031s^X^@^@^@1993-11-22 00:00:00.0000(^D^@^@^@s^F^@^@^@Dollars^F^@^@
^@D-Markf^R1.7037999629974365s^X^@^@^@1993-11-22 00:00:00.0000(^D^@^@^@s^F^@^@^@

Dollars^D^@^@^@Liraf^F1680.0s^X^@^@^@1993-11-22 00:00:00.0000(^D^@^@^@s^F^@^@^@D
ollars^C^@^@^@Yenf^R108.43000030517578s^X^@^@^@1993-11-22 00:00:00.0000(^D^@^@^@
s^F^@^@^@Dollars^G^@^@^@Guilderf^R1.9114999771118164s^X^@^@^@1993-11-22 00:00:00
.0000(^D^@^@^@s^F^@^@^@Dollars^F^@^@^@SFrancf^R1.4945000410079956s^X^@^@^@1993-1
1-22 00:00:00.0000(^D^@^@^@s^F^@^@^@Dollars^E^@^@^@Poundf^S0.67773997783660889s
^X^@^@^@1993-11-22 00:00:00.0000(^D^@^@^@s^E^@^@^@Pounds^F^@^@^@FFrancf^R8.73400
02059936523s^X^@^@^@1993-11-22 00:00:00.0000(^D^@^@^@s^E^@^@^@Pounds^C^@^@^@Yenf
^R159.99000549316406s^X^@^@^@1993-11-22 00:00:00.0000(^D^@^@^@s^C^@^@^@Yens^E^@
^@^@Poundf^U0.0062500000931322575s^X^@^@^@1993-11-22 00:00:00.0000(^D^@^@^@s^F^@
^@^@CdnDlrs^F^@^@^@Dollarf^R0.7534099817276001s^X^@^@^@1993-11-22 00:00:00.0000(
^D^@^@^@s^F^@^@^@CdnDlrs^F^@^@^@FFrancf^R4.4597001075744629s^X^@^@^@1993-11-22 0
0:00:00.0000t
^@^@^@insert_sqls\^@^@^@insert into cross_rate (from_currency, to_currency, conv
_rate, update_date) values (?,?,?,?)0

The resulting serialized dump file yielded the correct corresponding dictionary when de-serialized:

```
>>> import marshal
>>> data = marshal.load(open('cross_rate.1.dump', 'r'))
>>> data
{'insert_sql': 'insert into cross_rate (from_currency, to_currency, conv_rate, update_date) values (?,?,?,?)', 'data':
[('Dollar', 'CdnDlr', 1.327299952507019, '1993-11-22 00:00:00.0000'), ('Dollar', 'FFranc', 5.9193000793457031, '1993-
11-22 00:00:00.0000'), ('Dollar', 'D-Mark', 1.7037999629974365, '1993-11-22 00:00:00.0000'), ('Dollar', 'Lira', 1680.0,
'1993-11-22 00:00:00.0000'), ('Dollar', 'Yen', 108.43000030517578, '1993-11-22 00:00:00.0000'), ('Dollar', 'Guilder',
1.9114999771118164, '1993-11-22 00:00:00.0000'), ('Dollar', 'SFranc', 1.4945000410079956, '1993-11-22
00:00:00.0000'), ('Dollar', 'Pound', 0.67773997783660889, '1993-11-22 00:00:00.0000'), ('Pound', 'FFranc',
8.7340002059936523, '1993-11-22 00:00:00.0000'), ('Pound', 'Yen', 159.99000549316406, '1993-11-22 00:00:00.0000'),
('Yen', 'Pound', 0.0062500000931322575, '1993-11-22 00:00:00.0000'), ('CdnDlr', 'Dollar', 0.7534099817276001, '1993-
11-22 00:00:00.0000'), ('CdnDlr', 'FFranc', 4.4597001075744629, '1993-11-22 00:00:00.0000')]}
```

The second evaluation test was performed on a research database containing more than 50 tables of which 48 million records in 20 tables were chosen for backup. This task took about 22 minutes to complete on a 2GHz G5 PowerPC system with 2GB onboard RAM, with 104 dump files totaling 9.54 gigabytes. For comparison, the 104 dump files were loaded into a newly created database (this task took about 9 hours and 10 minutes) on and re-backup using gbak on the same system. The comparative backup using *gbak* took about 26 minutes and restoring the resulting *gbak*-backup-ed file took about 51 minutes. The results for this backup/restore comparison is summarized in Table 1 below for clarity.

|  | Using gbak utility | This study |
| --- | --- | --- |
| Time needed to backup 48 million records in 20 tables | 9 hours 10 minutes | 22 minutes |
| Time needed to restore 48 million records in 20 tables | 51 minutes | 26 minutes |

Table 1: Comparison of the length of time needed to backup and restore a test set of 48 million records in 20 tables of a Firebird database using gbak and the proposed method in this study.

## 4. Discussion

The standard Firebird database backup utility, *gbak*, backs up database into a compressed textual format (XDR), in contrast to MySQL's *mysqldump* utility which backs up database into a plain text format of SQL insert statements. There are advantages and disadvantages in each method. Firebird's method allowed for optimization of the indexes during the backup and restore process while MySQL's method enabled single table backup which tends to be more resilient to data corruption in the backup file. The main disadvantage of Firebird's method is its requirement for an uncorrupted database structure to be backed up successfully while MySQL's method does not allow certain optimizations of the database, such as index balancing, during the backup/restore process. Despite the advantages and disadvantages of each method, there are practical situations where one method might be preferred over another.

This paper presented a simple utility based on Python/kinterbasdb setup for table-level backup and restore in a text-coded format, using *marshal* module in Python Standard Library. The evaluation tests demonstrated that this utility performed with reasonable efficiency of backing up about 2 to 2.2 million records in a minute on a production system which is comparable to that of gbak. The evaluation results illustrated that restoring the database using the proposed utility is significantly slower than that of gbak, this might be use to the serialized use of the restore utility. That is, the dump files are restored one at a time. However, several instances of the restore utility (fb_load.py) could be executed in parallel (restoring different sets of dump files in each instance) and might significantly improve performance, especially on a multi-CPU/multi-core platform with Firebird classic version which allows for multiple instance of Firebird to be concurrently active.

As this utility is based on *marshal* module, Firebird tables containing un-marshallable objects, such as timestamp or date, cannot be dumped without conversion as shown in the evaluation test where *update_date* attribute (which is date type) had to be casted into 25-character attribute before serialization is possible. Hence, there are some limitations of this proposed method posed by data type mismatch between Firebird data types and marshallable data types. Despite the limitations, it might be possible to circumvent the limitation by type-casting as demonstrated in the evaluation test. Future improvements into the increasing range of serializable data types by *marshal* module may assist in reducing this limitation observed. At the same time, this utility had not been tested with BLOB objects in spite of kinterbasdb's capabilities to handle BLOBs using file-like input/output streams (please refer to kinterbasdb's usage guide for more information). Therefore, we believe that BLOB object handling may be currently possible with modifications and may be included in future versions.

Comparing with gbak, restoring a database using this utility does not result in an optimized set of indexes. However, this could be easily remedied as re-creating table indexes in Firebird is simply a 2-step process of dropping (deleting) a current index (using DROP INDEX SQL statement) and re-creating the same index (using CREATE INDEX SQL statement). These index re-creation could easily be made into an SQL script for routine database maintenance; hence, is not considered as an important feature for this utility.

In summary, we had presented a simple text-formatted backup and loading utility to perform table-by-table backup of a Firebird database for serialize-able or type-cast-able to serialize-able Firebird data types.